\documentclass[aps,prb,twocolumn,superscriptaddress,floatfix]{revtex4}
\usepackage{epsfig}
\usepackage{amsmath}
\usepackage{amsfonts}
\usepackage{amssymb}
\usepackage{mathptmx}
\usepackage{eucal}

\newcommand{\wo}{ \widetilde{\omega} }
\newcommand{\la}{ \left\langle }
\newcommand{\ra}{ \right\rangle }

\begin{document}
\title{Characteristics of the electric field accompanying a
longitudinal acoustic wave in a metal. Anomaly in the superconducting
phase}

\author{Yu.\ A.\ Avramenko} \author{E.\ V.\ Bezuglyi} \author{N.\ G.\ Burma} \author{I.\ G.\
Kolobov} \author{V.\ D.\ Fil'} \author{O.\ A.\ Shevchenko}
\affiliation{B. Verkin Institute for Low Temperature Physics and Engineering,
National Academy of Sciences of Ukraine, pr.\ Lenina 47, 61103 Kharkov,
Ukraine}

\author{V.\ M.\ Gokhfeld}

\affiliation{A. A. Galkin Donetsk Physicotechnical Institute, ul.\ R.
Lyuksemburg 72, 83114 Donetsk, Ukraine}

\begin{abstract}

The temperature dependence of the amplitude and phase of the electric
potential arising at a plane boundary of a conductor when a longitudinal
acoustic wave is incident normally on it is investigated theoretically and
experimentally. The surface potential is formed by two contributions, one of
which is spatially periodic inside the sample, with the period of the
acoustic field; the second is aperiodic and arises as a result of an
additional nonuniformity of the electron distribution in a surface layer of
the metal. In the nonlocal region the second contribution is dominant. The
phases of these contributions are shifted by approximately $\pi /2$. For
metals in the normal state the experiment is in qualitative agreement with
the theory. The superconducting transition is accompanied by catastrophically
rapid vanishing of the electric potential, in sharp contrast to the
theoretical estimates, which predict behavior similar to the BCS dependence
of the attenuation coefficient for a longitudinal sound.

\end{abstract}

\maketitle

\section*{1. INTRODUCTION}

When an elastic wave propagates in a metal, a perturbation of its
electron subsystem occurs which compensates the tendencies of the ionic
displacements to disrupt the charge neutrality or the balance between
the ion and electron currents. At high temperatures or in dirty samples
(the so-called local limit, determined by the condition $ql \ll 1$,
where $q$ is the acoustic wave vector and $l$ is the electron mean free
path) the electron collision frequency is high enough to maintain local
equilibrium at any point of the deformed lattice. In the limit $ql \ll
s/v_{F}$ ($s$ and $v_{F}$ are the sound velocity and Fermi velocity)
the electric field arising in the metal has a purely inertial nature
--- the acoustic analog of the Stewart--Tolman effect.

In the opposite limiting case ($ql \gg 1$, nonlocal limit) the collisions can
no longer maintain local equilibrium, and additional electric fields are
produced in the metal, bringing about the required adjustment of the
electrons to the moving lattice. In zero magnetic field the polarization of
the electric field of an elastic wave propagating in an isotropic metal or
along a high-symmetry direction is parallel to the ionic displacements. When
a magnetic field is turned on, the Lorentz force exerted on the electrons
moving with the lattice gives rise to a Hall component of the alternating
electromagnetic field, the existence of which is no longer dependent on
strict requirements on the purity or temperature of the sample.

The opposite process also occurs: the generation of acoustic vibrations when
a polarized electromagnetic field is excited in the metal in a suitable way.
On the whole, these phenomena --- the acoustoelectric and electroacoustic
transformations --- are well known and widely used both in scientific
experiments and for contactless excitation of sound in technical applications
(see, e.g., the review\cite{1}).

However, as far as the authors are aware, the experimental studies have
always been done for electromagnetic fields in which the electric field
vector was orthogonal to the wave vector of the elastic wave.\cite{1a} The
latter is apparently due to the fundamental differences in the behavior of
the fields of transverse and longitudinal polarization at the interface
between the metal and free space (or an insulating medium). For normal
incidence of a transverse elastic mode on the interface the electromagnetic
field in the metal is naturally ``matched'' with the electromagnetic field in
the vacuum (insulator). In other words, in this geometry an electromagnetic
wave is radiated from the metal; this wave can easily be detected by a
suitably oriented antenna of the flat-coil type. This same antenna can also
be used to generate a transverse elastic wave in the metal.

An electromagnetic wave with electric field polarized along the wave vector
does not exist. Therefore, in the case of normal incidence of a longitudinal
acoustic wave on the interface, electromagnetic radiation from the sample is
absent in principle. Nevertheless, it is possible to detect the electric
field ${\bf E}$ due to a longitudinal acoustic wave. Since ${\rm div}{\bf
E}\neq 0$ in a longitudinal wave, uncompensated charge and the accompanying
potential appear both in the bulk of the metal and on its surface.\cite{4}
This potential can be detected by a voltmeter of suitable sensitivity. One of
the authors (V.M.G.) long ago had analyzed a method of linear electroacoustic
conversion with the use of a contactless capacitive driver.\cite{5} Here,
when an alternating voltage is applied to the exciting plate of the
capacitor, an electric charge is induced beneath it on the surface of the
metal, and the nonlocal interaction of this charge with the lattice excites a
longitudinal elastic wave of the same frequency. A detailed experimental
check of Ref.\ \onlinecite{5} has not been made, possibly because of the
rather stringent requirements on the purity of the metal. The limiting case
of a capacitive driver (receiver) is a galvanic contact, which is what was
mainly used in the present study. Finally, the presence of the contact on the
interface alters the boundary conditions for the elastic deformations;
however, it turns out that the latter leads to a slight increase in the
conversion efficiency and in general relaxes the requirements on the purity
of the material under study. For this reason it is possible to carry out an
experiment on practically all reasonably pure metals provided that some
relatively simple requirements are met as to the nonlocality of the
interaction of the electrons with the elastic field ($ql>1)$.

The structure of this paper is as follows. In Sec.~2 a theory of the
acoustoelectric conversion is given for the simplest case --- a conducting
half space (one-dimensional problem) in the approximation of an isotropic
dispersion relation for the carriers and their ``specular'' reflection by the
interface. The electric potential accompanying an elastic wave in the metal
is a sum of two contributions. Besides the well-known forced solution, which
repeats the profile of the elastic deformations, in the nonlocal limit there
is also a substantial quasiperiodic term due to the perturbation of the
ballistic motion of the electrons by the surface of the sample.

In Sec.\ 3 we present the results of experimental studies of the temperature
dependence of the aforementioned electric potential in metals of different
degrees of purity (Ga, W, Al) in the normal state. On the whole, the
temperature dependences of the amplitude and phase of the potential show an
acceptable qualitative agreement with the theoretical estimates; this can be
regarded as experimental confirmation of the theoretical ideas proposed here.

In Sec.\ 4 we discuss the evolution of the measured electric potential at the
superconducting transition. Contrary to the expectation of a rather slow
decrease in the recorded value below $T_{C}$ (like the behavior of the
longitudinal sound attenuation coefficient), a much sharper decrease in the
amplitude of the potential is observed.

\section*{2. THEORY OF ACOUSTOELECTRIC CONVERSION IN THE CASE OF NORMAL
INCIDENCE OF A SOUND WAVE ON AN INTERFACE}

Let us consider a metallic half space $x \geq 0$. Suppose that a longitudinal
elastic wave $u_{x} = u_0\exp(-i\omega t-iqx)$ comes in from the interior of
the sample and is reflected from the sample boundary with a reflection
coefficient $R$ (we neglect the sound attenuation). Near the boundary the
displacement field and the deformation field are formed as a result of the
interference of the incident and reflected waves, independently of the regime
(pulsed or continuous). The resultant field is expressed in terms of the
amplitudes of the displacement $u(0)$ and deformation $u'(0)$ at the
interface. In particular, for the deformation field that we will be
interested in below we have
\begin{eqnarray}
\frac{du_x(x)}{dx} \equiv u_{xx}(x) = u'(0) \cos qx - qu(0) \sin qx.
\end{eqnarray}

The relation between $u(0)$ and $u'(0)$ is determined by the boundary
conditions
\begin{eqnarray}
\frac{qu(0)}{iu'(0)} =\frac{1+R}{1-R} \equiv C.
\end{eqnarray}

For a contact with an elastically uniform, nonattenuating medium the
parameter $C$ is equal to the ratio of the acoustic impedances of the
metal and of the medium in contact with it (in particular, $C = \infty
$ for a contact with the vacuum). In the case of an attenuating medium
the parameter $C$ can be complex-valued.

The deformation-induced deviation $\psi (x,v_{x})\partial f_0/\partial
\varepsilon $ of the electron distribution function from its equilibrium
value $f_0$ is described by the kinetic equation\cite{6} (we neglect the
inertial field):
\begin{eqnarray}
v_x \frac{\partial}{\partial x}(\psi -e\varphi) -i \wo \psi = i\omega
\Lambda_{xx} u_{xx}.
\end{eqnarray}
Here $\wo\equiv \omega +i/\tau $ ($\omega $ is the angular frequency, and
$\tau $ is the electronic relaxation time), $\Lambda _{xx}$ is the
corresponding component of the deformation potential tensor $\Lambda _{ik} =
\lambda _{ik}-\langle \lambda _{ik}\rangle /\langle 1\rangle $, $\lambda
_{ik}$ is the ``bare'' deformation potential, $\varphi (x)$ is the electric
potential in the field of the elastic wave, $v_{x}$ is the $x$ component of
the Fermi velocity $v_{F}$ of the electron, and the angle brackets denote
averaging over the Fermi surface with a weight $v_{F}^{-1}$:
\begin{eqnarray*}
\langle A\rangle \equiv \frac{2}{h^3}\int \frac{A\, dS}{v_F}.
\end{eqnarray*}

It is important to note that the quantity $e\varphi (x)$ appearing in Eq.\
(3) is actually the total electrochemical potential of the electrons,
including, in addition to the ``true'' electric potential, the change of the
chemical potential $u_{xx}\langle \lambda _{xx}\rangle /\langle 1\rangle $
due to the change of the electron spectrum in the elastic deformation
field.\cite{6} The difference of the electrochemical potentials at different
points of the sample is a source of real emf that can be registered by a
voltmeter, and it, of course, vanishes at equilibrium, e.g., in the limit of
a static deformation ($\omega \rightarrow 0$). At the same time, the gradient
of the ``true'' electric potential, which compensates the deformation
contribution to the electrochemical potential and brings about an adjustment
of the electron density to the spatial variations of the ion density is
always nonzero in a nonuniformly deformed sample (physically this effect is
analogous to a contact potential difference). The presence of this potential
in a nonuniformly deformed metal (including in the elastic field of a sound
wave) gives rise to uncompensated charges with a density $\delta n =
r_{D}^{2}\langle \lambda _{xx}\rangle \nabla ^{2}u_{xx}\sim (qr_{D})^{2}qun$,
where $n$ is the total electron density, $q$ is the characteristic wave
number of the deformation, and $r_{D}$ is the screening radius of the
longitudinal field ($r_{D}^{-2} = 4\pi e^{2}\langle 1\rangle )$, which in
``good'' metals is of the order of the lattice constant. Such a ``charge
density wave'' accompanying the propagation of a sound wave in the general
case contains a nonequilibrium contribution due to the disruption of the
spatial uniformity of the electrochemical potential $e\varphi (x)$; this
contribution is proportional to the frequency of the sound (see below) but in
the majority of cases it is small (${\sim }s/v_{F})$ compared to the
``adiabatic'' component mentioned above. To avoid misunderstandings we should
say that, because of the small value of the uncompensated charge ($\delta n
\ll n)$ the potential $\varphi (x)$ can, of course, be calculated from the
condition of electrical neutrality $\langle \psi \rangle  = 0$, where
$\langle \psi \rangle $ has the meaning of a nonequilibrium admixture to the
charge density which is ``adiabatically'' modulated by the elastic field.

Far from the boundary ($x \gg l)$ one can assume that the functions $\psi
(x)$ and $\varphi (x)$ are periodic, with the same spatial period as
$u(x)$,\cite{6} and in this case Eq.~(3) reduces to an algebraic equation.
Near the interface (at distances $x \leq l$ from it) the electron
distribution differs substantially from periodic, and the problem of finding
$\psi (x)$ and $\varphi (x)$ is complicated. It can be solved relatively
simply by the Fourier method if the so-called specular boundary condition is
imposed on the function $\psi (x)$:
\begin{eqnarray}
\psi(+0,v_x)=\psi(+0,-v_x).
\end{eqnarray}

We emphasize that the condition of ``specular'' reflection from the
moving boundary can be written in the form (4) only in a comoving
reference frame, in which there is no current through the boundary
($\langle v_{x}\psi (0,v_{x})\rangle  = 0$), and the scattering of the
electrons is elastic. Since the potential measurement is actually done
at a moving boundary, all of the calculations below pertain to the
comoving system, in which the form (3) of the linearized kinetic
equation is preserved.

Let us continue the functions $u_{xx}(x)$ and $\varphi (x)$ evenly onto
the semiaxis $x < 0$ (this will be denoted below by a superscript $S$).
Then from the form of Eq.\ (3) (in view of the evenness of $\Lambda
_{xx}(v_{x})$ it is symmetric with respect to a simultaneous change of
the signs of $x$ and $v_{x})$ and condition (4) it follows that the
unknown function $\psi (x)$ should be continued to $x<0$ without a
discontinuity. As a result, its transform is equal to
\begin{eqnarray}
\psi_k =\frac{kv_x(\varphi^S)_k -\omega\Lambda_{xx}(u^S_{xx})_k}{kv_x - \wo}.
\end{eqnarray}
The Fourier transform of the electric potential $(\varphi ^{S})_{k}$ of
interest is found from the condition of electrical neutrality of the
metal, $\langle \psi _{k}\rangle  = 0$:
\begin{eqnarray}
(\varphi^S)_k = (u^S_{xx})_k\la \frac{\omega\Lambda_{xx}}{kv_x-\wo}\ra \la
\frac{kv_x}{kv_x-\wo} \ra^{-1} \equiv (u^S_{xx})_k R_k,
\end{eqnarray}
where $(u^S_{xx})_k$ is the transform of the deformations (1),
\begin{eqnarray}
(u^S_{xx})_k = q \left[ \frac{qu(0) -iu'(0)}{k^2 -(q+i0)^2} +\frac{qu(0)
+iu'(0)}{k^2 -(q-i0)^2}\right].
\end{eqnarray}

Integration over the Fermi surface in (6) leads to branch points $k = \pm K$
($K = \tilde{\omega }/v_{F})$ of the kinetic coefficient $R_{k}$; these
singularities are in addition to the ``acoustic'' poles $k = \pm (q\pm i0)$
of expression (7). Inversing the Fourier transform, we find that in addition
to the periodic forced solution, which is determined by the poles (the $q$
contribution), the function $\varphi (x)$ also contains an aperiodic term,
whose amplitude is a complicated function of $x$ and whose phase (${\sim}K
x)$ is determined by the Fermi velocity ($K$ contribution).

The asymptotics of the solutions of this type at large depths have been
discussed previously on more than one occasion in the analysis of the
propagation of electron quasiwaves in a metal.\cite{5,7,8} In the present
paper we will be interested in the relationship of the $K$ and $q$
contributions to the electric potential $\varphi (0)$ measured at the
boundary of the metal and their respective temperature dependences.

Let us see what the solution (6) gives in the simple case of a quadratic
isotropic dispersion relation of the charge carriers, $\varepsilon  =
p^{2}/2m$, when the deformation potential can be represented in the
form\cite{6}
\begin{eqnarray}
\Lambda_{xx} =L\bigl(3v_x^2/v_F^2 -1\bigr).
\end{eqnarray}
In this case the kinetic factor in (6) is equal to
\begin{align}
&R_k =\frac{\omega L}{\wo} \Phi(z), \quad \Phi(z) = \frac{3}{z^2} - \frac{\ln
\frac{1-z}{1+z}}{2z +\ln \frac{1-z}{1+z}}, \nonumber
\\
&z = \frac{kv_F}{\wo} = \frac{kl}{\omega \tau+i}.
\end{align}
To find the coefficient of proportionality between $u_{xx}(x)$ [see Eq.\ (1)]
and the $q$ contribution to $e\varphi (x)$, it is sufficient to substitute $k
= q$ in (9); thus, in the $x$ representation we obtain
\begin{eqnarray}
e\varphi_q(x) =R_q u_{xx}(x),\quad R_q =L\frac{s}{v_F}a\Phi(a),\quad a \equiv
\frac{v_F}{s}\frac{\omega\tau}{\omega\tau+i}.
\end{eqnarray}

Let us now evaluate the $K$ contribution. According to what we have said, it
is given by the integral
\begin{eqnarray}
e\varphi_K(x) =\int_C \frac{dk}{2\pi} (U^S_{xx})_k R_k \exp(ikx)
\end{eqnarray}
over a contour $C$ passing along the edges of the branch cut $k =
y\widetilde{\omega }/v_{F}$ ($1 \leq y<\infty )$ in the complex $k$ plane.
Using formulas (6), (7), and (9), we can write this integral in the form
\begin{align}
e\varphi_K(x) &= i\omega u(0) \frac{L}{v_F} \int_1^\infty \!\!\!\!\!dy
\frac{ya^2}{a^2-y^2} \Bigl[\Bigl(y + \frac{1}{2} \ln \frac{y-1}{y+1}\Bigr)^2
\!\!\!+ \frac{\pi^2}{4}\Bigr]^{-1} \nonumber
\\
&\times \exp\Bigl(\frac{iy\wo x}{v_F}\Bigr) \equiv i\omega u(0)\frac{L}{v_F}
J(a,x).
\end{align}

We note that the $K$ contribution at any $x$ is proportional to the
displacement of the surface, $u(0)$, while the $q$ contribution is
proportional to the local deformation $u'(x)$. This means that upon a
variation of the surface potential by means of a capacitor with a vacuum
(gas-filled) gap --- an elastically free surface --- we can record only the
$K$ contribution. In the general case, by adding (10) and (12) together we
obtain for the potential measured on the surface:
\begin{eqnarray}
\varphi(0)=\varphi_q(0)+\varphi_K(0) = \frac{iL}{ev_F}\Bigl(\frac{2I}{\rho
s}\Bigr)^{1/2}\Bigl[J(a,0)-\frac{a\Phi(a)}{C}\Bigr],
\end{eqnarray}
where $I$ is the energy flux in the sound beam, and $\rho $ is the density of
the conducting medium.

\begin{figure}
\centerline{\epsfxsize=8cm\epsffile{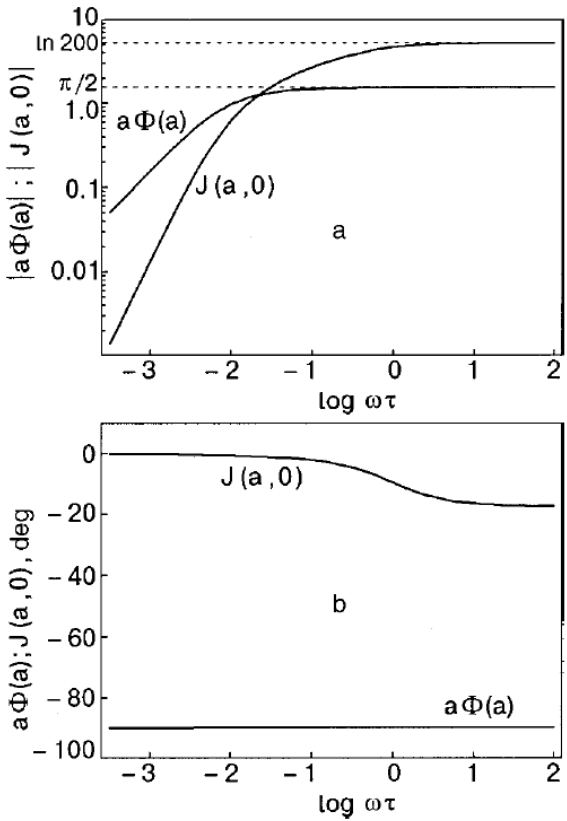}} \vspace{-5mm}
\caption{Theoretical dependence of the amplitude (a) and phase (b) of the
periodic ($\Phi )$ and aperiodic ($J)$ contributions to the surface electric
potential, as functions of the scattering parameter. } \label{f1}
\vspace{-5mm}
\end{figure}

Let us estimate the expected value of the effect. The coefficient in formula
(13) at $L\sim 10$ eV and $I\sim 10$ W/cm$^{2}$ has a value close to ${\simeq
}1$ $\mu$V. The dependence of the functions $|a\Phi (a)|$ and $|J(a,0)|$ on
the parameter $\omega \tau $ is presented in Fig.\ \ref{f1}. In the nonlocal
region of frequencies and temperatures ($ql>1)$ the first of these rapidly
approaches its limiting value $|a\Phi (a)| = \pi /2$, whereas the second
varies as $\ln|a|$; for $\omega \tau  \gg 1$ its limiting value
$|J(v_{F}/s,0)|\approx \ln(v_{F}/s)$, which is practically always greater
than $\pi /2$ (the curves in Fig.\ \ref{f1} were constructed for the
parameter value $v_{F}/s = 200$ ($\ln v_{F}/s = 5.3$), which is typical of
gallium). Thus the expected value of the effect is at the level of a few
microvolts, i.e., quite amenable to measurement.

In Fig.\ \ref{f1} we see that for $C\sim 1$ the contributions under
discussion become equal at $ql\sim 1$. The phases of these contributions,
however, are shifted by $\pi /2$, and therefore on the curves of the
temperature dependence of $\varphi (T)$ one should expect an appreciable
change in the phase of the signal in the temperature region where the mean
free path becomes comparable to the wavelength of the sound.

Of course, the specular boundary condition used does not fully correspond to
the experimental situation; however, it is known that taking more realistic
boundary conditions (the diffuseness of the boundary) into account in
nonlocal problems of acoustoelectronics complicates the calculations
significantly without leading to any substantial changes in the
results.\cite{9} It can be hoped that in the given case the behavior of
$\varphi (0)$ will correspond qualitatively to the behavior that follows from
(13).

At the parameters values used, the amplitude of the adiabatic component of
the charge density wave accompanying the propagation of a sound wave in a
metal is of the order of 10$^{8}$ cm$^{-3}$. Let us now estimate the
contribution of the nonequilibrium potential $\varphi (x)$ to the amplitude
of the wave of uncompensated charge density. In the general case one can find
its value by solving Eq.\ (5) jointly with Poisson's equation $k^{2}\varphi
_{k} = 4\pi e\langle \psi _{k}\rangle $. As a result, for the nonequilibrium
admixture to the charge density we obtain
\begin{eqnarray}
\la \psi_k\ra =-(u^S_{xx})_k \la\frac {\omega\Lambda_xx}{kv_x - \wo}\ra
\biggl( 1+\frac{4\pi e^2}{k^2}\la\frac {kv_x}{kv_x - \wo}\ra\biggr)^{-1}.
\end{eqnarray}

The final answer depends on the relationship between the squares of the
inverse screening radius and the actual wave number. For the periodic
component of the charge density wave ($qr_{D} \ll 1)$ the 1 in the
denominator of (14) can be neglected at any distance from the surface of the
metal, i.e., the answer reduces to a double differentiation of the potential
$\varphi _{q}$ found previously from the condition of electrical neutrality.
As a result, the amplitude of this component turns out to be small (${\sim
}s/v_{F})$ in comparison with the adiabatic contribution. To estimate the $K$
contribution at distances $x \gg r_{D}$ from the surface this approximation
is again applicable, since in this case the convergence of the integrand in
(12) (which acquires an additional factor of $k^{2}$ after the double
differentiation with respect to $x$) at large $k$ is provided by the
exponential factor, and the unity in the denominator of (14) can again be
neglected. This allows us to obtain the following estimate for the
uncompensated charge of the $K$ component at distances from the surface which
are small compared to the mean free path and the wavelength of the sound:
\begin{eqnarray}
\delta n(x) -i\omega u(0) \frac{L}{v_f} \frac{q^2}{4\pi e^2}
\ln\frac{v_F}{x\wo}.
\end{eqnarray}
Thus the uncompensated charge of the $K$ component increases logarithmically
as $x \rightarrow 0$, reaching a maximum value in the surface region $x  \leq
r_{D}$. The density of this surface charge can be estimated to logarithmic
accuracy by employing a cutoff at $k\sim r_{D}^{-1}$ in the total denominator
of expression (14) and making the substitution $x\rightarrow r_{D}$ in Eq.\
(15); this leads to a value $\delta n\sim 10^{8}$ cm$^{-3}$, which is
comparable to the amplitude of the adiabatic component.

We conclude this Section by verifying the applicability of the reciprocity
theorem for the ``exotic'' mechanism of acoustoelectric coupling under
consideration. In this case the acoustoelectric converter would be of the
electrostatic type, in which the ponderomotive forces are due to the
interaction of the electric charges. For converters of this type the
electromechanical reciprocity theorem can be formulated as\cite{10}
\begin{eqnarray}
\frac{\partial F}{\partial g} = \frac {\partial \varphi}{\partial u},
\end{eqnarray}
where $F$ is the density of the mechanical force, which is causally
related to the uncompensated charge density $g$, while the remaining
components are the same as those already used. The relation between the
electric field and the periodic elastic displacement is written in the
form ${\bf u} = B{\bf E}$. Since $F = \rho s^{2}d^{2}u/dx^{2}$ and $g =
{\rm div}{\bf E}/4\pi $ (since we are seeking an harmonic component of the
displacements, differentiation with respect to $x$ reduces to
multiplication by $q$), and, using formula (12) for a free surface ($C
= \infty )$, we find $B\approx L\ln(v_{F}/s)/(4\pi \rho sev_{F})$,
which agrees exactly with the result obtained in Ref.\ \onlinecite{5}.

\section*{3. EXPERIMENTAL STUDY OF THE SOUND-GENERATED SURFACE
POTENTIAL IN CONDUCTORS FOUND IN THE NORMAL STATE}

\begin{figure}[tb]
\centerline{\epsfxsize=5cm\epsffile{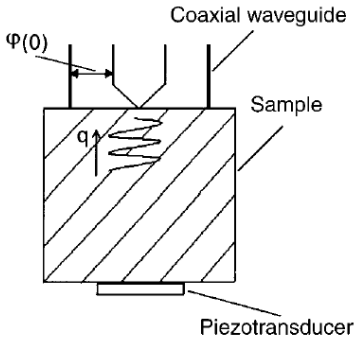}}
\vspace{-5mm} \caption{Diagram of the experiment. } \label{f2} \vspace{-5mm}
\end{figure}

The basic scheme of the experiment is illustrated in Fig.\ \ref{f2}. The
sample under study is used to short a coaxial feeder, the outer conductor of
which is in contact with the sample outside the region of incidence of the
sound beam. The inner, spring-loaded conductor is a fragment of a hemisphere
of rather large radius (${\sim }1$ cm). Thus the experiment measures the
amplitude and phase of the potential difference between the central part of
the ``hot'' spot created by a sound beam of diameter ${\sim }4$ mm and the
remote parts of the surface of the sample at zero potential. In some
experiments the galvanic contact of the inner conductor of the coaxial feeder
with the sample was replaced by a capacitive coupling ($C\sim 5$ pF), or in
some cases a flat (planar) coil displaced relative to the axis of symmetry of
the sound beam was used to detect the magnetic field of the currents
spreading out from the ``hot'' spot (this will be referred to below as an
asymmetric coil).

An rf oscillator ($\omega /2\pi \sim 55$ MHz) and a lithium niobate
piezotransducer were used, giving a maximum acoustic power per pulse of up to
50--100 W/cm$^{2}$. The pulse duration (${\sim }5\times 10^{-7}$ s) and the
pulse repetition rate (${\sim }17$ Hz) were chosen so that the heating of the
sample at a temperature $T\sim 1$ K did not exceed $(2$--$3)\cdot 10^{-2}$ K.

{\em Gallium.} Single-crystal samples of high-purity gallium were used,
ensuring a parameter value $\omega \tau \sim 5$ at the working frequency in
the region of impurity scattering.

When the potential was registered using the galvanic contact, in addition to
the signal coincident in time of arrival with the acoustic pulse, an
electron-sound signal, passing through the sample at the Fermi velocity, was
also observed (for temporal separation of the latter from the signal
appearing at the time of the probe pulse, a germanium delay line, not shown
in Fig.\ \ref{f2}, was used). Analysis of the nature of the electron sound is
proposed as the subject of a separate study; in this paper we discuss only
the electric potential arising on the surface of the sample at the time of
arrival of the sound pulse.

The maximum response at a fixed excitation power could be attained only for
freshly ground surfaces; prolonged storage of a sample led to a falloff of
the signal amplitude, apparently because of diffusion of impurities into the
subsurface region. At $T = 1.5$ K the value of $|\varphi (0)|$ at the maximum
excitation power reached 30--50 $\mu$V. The characteristic value of the Fermi
velocity on the main sheets of the Fermi surface of gallium is $v_{F}\approx
7\times 10^{7}$ cm/s,\cite{11} and therefore in accordance with the estimate
(13) a large value of the deformation potential ($L\approx 20$ eV) is
required in order to obtain such values of $\varphi (0)$. It is also possible
that the main contribution to the effect is given by sheets of the Fermi
surface on which the Fermi velocity is low, which are known to exist in
gallium.\cite{11}

\begin{figure}[tb]
\centerline{\epsfxsize=8cm\epsffile{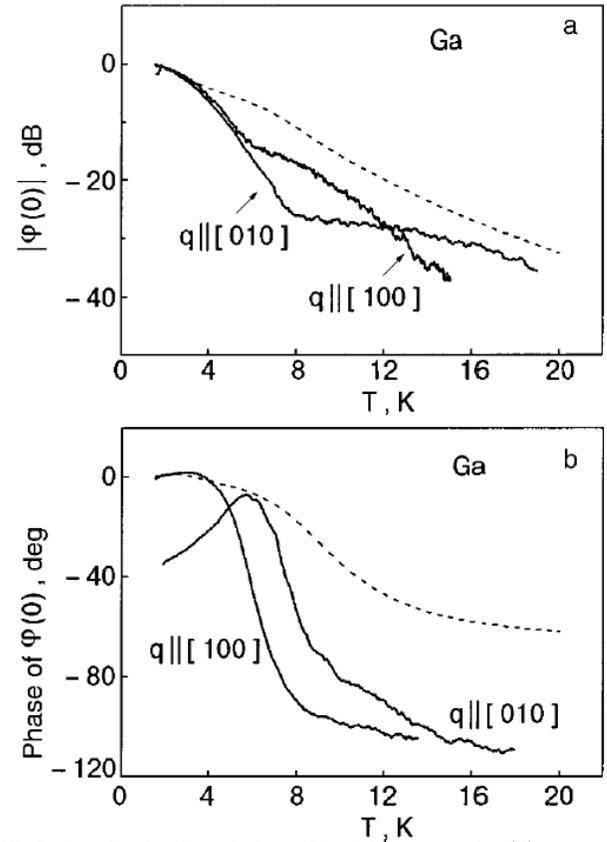}}
\vspace{-5mm} \caption{Amplitude (a) and phase (b) of the potential $\varphi
(0)$ measured with a galvanic contact on Ga in the propagation of sound along
different crystallographic directions (solid curves) at a frequency of 55
MHz. The dashed curve shows the result of a calculation according to Eq.\
(13) for $C = 1$, $(\omega \tau )^{-1} = 0.2+0.05T^{2}+0.15T^{3}$, $v_{F}/s =
200$. } \label{f3} \vspace{-7mm}
\end{figure}

Typical curves of the temperature dependence of the amplitude and phase of
the potential measured with the use of a galvanic contact are presented in
Fig.\ \ref{f3}. The data for $|\varphi (0)|$ are corrected for the change in
sound attenuation in the sample. The resultant phase of the signals is
determined not only by the phase of $\varphi (0)$ but also, and mainly, by
the acoustic delay. However, in the investigated temperature interval the
corrections due to the change in the sound velocity did not exceed a few
percent of the measured total variations of the phase.

For Ga the temperature dependence of the relaxation time is known quite
well,\cite{12} and therefore in Fig.\ \ref{f3} we also show the calculated
curves obtained at $C = 1$. The calculations were done on the assumption that
low-angle scattering is efficient $[(\omega \tau )^{-1} =
0.2+0.05T^{2}+0.15T^{3}]$, although this question requires separate
investigation. Without going into the details of the anisotropy of the
effect, which is clearly seen on the experimental curves, we will mention the
two circumstances that we think are essential.

1. As expected from the analysis given above, the phase of $\varphi (0)$
experiences significant (${\sim }\pi /2)$ variation in the temperature region
corresponding to the transition from the nonlocal to the local regime. On the
whole, the scale of the variations of the amplitude and phase of $\varphi
(0)$ agree with the calculation. There is justification for thinking that the
stated theoretical ideas about the nature of the origin of $\varphi (0)$ and
the decisive role of the $K$ contribution at large values of $ql$ are in
qualitative agreement with experiment.\cite{15a}

2. The calculation gives a much smoother variation of the amplitude and phase
of $\varphi (0)$ in the crossover region than is observed in experiment. The
kink in the temperature dependence of the amplitude, which coincides with the
center of the ``jump'' in phase, cannot be described using Eq.\ (13), even
when the model parameters are varied over wide limits. Apparently, the rate
of decrease of the $K$ contribution in the crossover region is substantially
higher than is predicted by Eq.\ (13). This circumstance is probably due to
the ``specularity'' and isotropicity of the dispersion relation which were
imposed on the model, and the solution for a more realistic case will be
closer to the observed behavior of $\varphi (0)$.

Let us also mention some other features of the behavior of the phase of
$\varphi (0)$. For $q\parallel [010]$ at $T<6$ K the phase of the signal
increases with increasing temperature (Fig.\ \ref{f3}b). This behavior occurs
only in that geometry, and it is most likely due to the broad flattening on
the Fermi surface of Ga in the ${\bf q}\cdot {\bf v} = 0$ region.\cite{13} It
is easy to see from the relations given above that in this case, for $\omega
\tau \sim 1$, the increase in the scattering leads to growth of the phase of
the $q$ contribution while having practically no effect on the $K$
contribution. The model calculation according to Eq.\ (13) with the actual
relative area of the flat part taken into account (${\sim }2$--$4\%$; Ref.\
\onlinecite{13}) gives a good description of both the scale of this effect
and its temperature dependence.

It follows from Fig.\ \ref{f3}b that the change in phase of the signal in the
crossover region is assuredly in excess of $\pi /2$. This is possible due to
the complex nature of the parameter $C$ in the case of reflection of sound
from a contact region of small size. If $C$ in Eq.\ (13) has the form $C =
C_0(1+i\beta )$ ($\beta >0$), then in the nonlocal parameter region, where
the contribution from the first term in (13) predominates, the influence of
$\beta $ is insignificant. However, on going to the local limit the phase of
the harmonic component turns out to be lower than the calculated value by an
amount $\arctan(\beta )$.

\begin{figure}[tb]
\centerline{\epsfxsize=8cm\epsffile{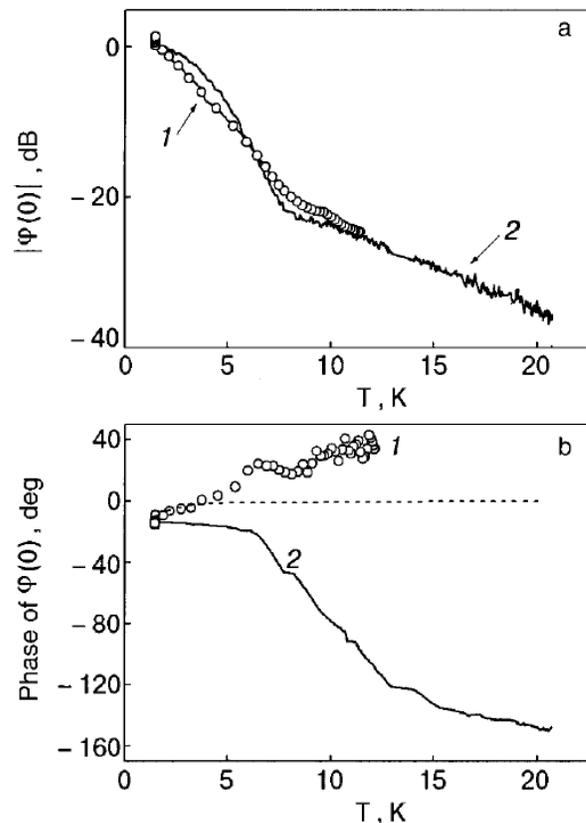}}\vspace{-5mm}
\caption{Amplitude (a) and phase (b) of the potential $\varphi (0)$ measured
on Ga with a capacitive sensor ({\em 1}) and an asymmetric coil ({\em 2})
(solid curves). The dashed curves show the results of a calculation according
to Eq.\ (12) (the parameters used in the calculation were the same as in
Fig.\ \ref{f3}). } \label{f4}\vspace{-5mm}
\end{figure}

In the proposed conception of the origin of $\varphi (0)$ the variation of
the phase of the recorded signal is due to the fact that the interface in the
contact region is elastically non-free ($u'(0)\neq 0$, i.e., $C^{-1}\neq 0)$.
To check this idea, we replaced the galvanic contact with a capacitor with a
vacuum gap (more precisely, a gap filled with a heat-exchange gas). The
result is shown in Fig.\ \ref{f4}. In view of the small value of the
capacitance (as we were striving to avoid edge effects, the electrode forming
the capacitor had a diameter of ${\sim }2$ mm) we were unable to make
measurements in the same temperature interval as in Fig.\ \ref{f3}.
Nevertheless, it is reliably established that in this case there is no
significant decrease of the phase of the signal. We assume that this result,
taken together with the data of Fig.\ \ref{f3}b, is unambiguous evidence in
support of the approach developed here. Moreover, relation (12) also predicts
a certain increase in the phase of the $K$ contribution with increasing
scattering even in the purely isotropic approximation; this is apparently
registered by the capacitive sensor, although on a larger scale (Fig.\
\ref{f4}b).

The presence of excess surface charge in the region of the ``hot'' spot
presupposes the existence of currents spreading out from the center of the
sample toward the periphery. To detect them we used a flat coil
asymmetrically shifted relative to the center of the sample. The plane of the
turns of this coil were oriented perpendicular to the interface and parallel
to the radial direction, the surface of the sample in the region of the
``hot'' spot was left elastically non-free. The results of this experiment
are also presented in Fig.\ \ref{f4}. The amplitude of the signal behaves
analogously to that shown in Fig.\ \ref{f3}a, while the phase of the signal
deviates even more to the high side of $\pi /2$. Most likely this is due to
the fact that the rf conductivity of the metal, being a complex quantity,
varies in phase on going to the local limit, thereby altering the phase of
the rf current as well, so that the total shift increases.

{\em Tungsten and aluminum.} In tungsten in the impurity-scattering region
the value of the parameter $\omega \tau $ is approximately equal to 1, while
in aluminum $\omega \tau \approx 0.3$. It should be noted, however, that our
estimates of the impurity scattering were obtained from a study of the bulk
characteristics and can be somewhat overestimated for the surface regions.
The values of $|\varphi (0)|$ measured in these metals turned out to be
substantially smaller than in gallium (${\sim }1$ $\mu$V). This is clearly
due to the small value of the deformation potential, since a slight (see
Fig.\ \ref{f1}) increase in the scattering (i.e., decrease in the parameter
$\omega \tau $ in comparison with gallium) should not lead to a substantial
decrease in $|\varphi (0)|$.

\begin{figure}
\centerline{\epsfxsize=8cm\epsffile{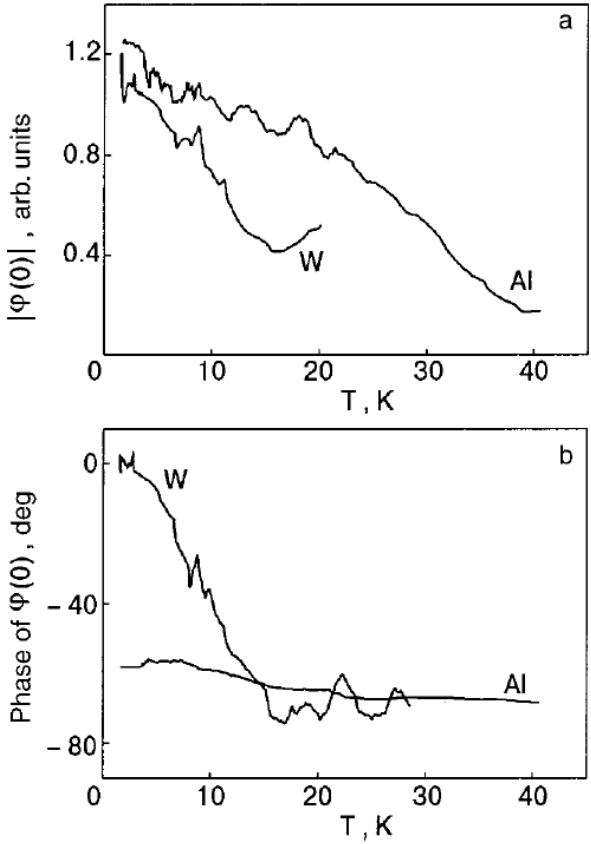}}\vspace{-5mm}
\caption[]{Amplitude (a) and phase (b) of the potential $\varphi (0)$
measured with a galvanic contact on W and Al. } \label{f5}\vspace{-5mm}
\end{figure}

Curves of the temperature dependence of the amplitude and phase of $\varphi
(0)$ for W and Al are shown in Fig.\ \ref{f5}. In spite of the small value of
the detected signal, the curves for tungsten qualitatively repeat the
behavior of $\varphi (0)$ in Ga, the most salient feature of which, in our
view, is the significant variation of the phase of the signal. In Al the
phase of the signal is practically unchanging. It can be supposed that, as in
Ga, the more rapid (in comparison with Eq.\ (13)) falloff of the $K$
contribution has led to its nearly complete suppression already at the
starting (impurity) value of $\omega \tau $.

\section*{4. INFLUENCE OF THE SUPERCONDUCTING TRANSITION ON $\varphi
(0)$}

Before turning to a description of the experimental results, let us briefly
present the theoretical scheme for estimating the possible behavior of
$\varphi (0)$ below $T_{C}$. It follows from the above discussion that the
existence of a measurable potential $\varphi (0)$ is due to the deformation
interaction of the electrons with the elastic field of a longitudinal wave.
In a superconductor only normal excitations interact with sound through the
deformation potential. The ``freezing out'' of the normal excitations leads
to the situation that for $T \ll T_{C}$ the potential $\varphi (0)$ is due
solely to the Stewart--Tolman inertial field, which we are neglecting in this
paper. Consequently, our problem consists in estimating the decay law of
$\varphi (0)$ below $T_{C}$.

In a superconductor the electromagnetic field is customarily described by
a gradient-invariant combination of $\Phi $ (the electrochemical
potential of the excitations) and ${\bf p}_{s}$ (the momentum of the
superconducting condensate) and the spatial and time derivatives of the
phase of the order parameter $\chi $ and the electromagnetic potentials
$\varphi $ and ${\bf A}$:
\begin{eqnarray}
\Phi =\frac{1}{2}\frac{\partial\chi}{\partial t} + e\varphi, \quad {\bf p}_s
= \frac{1}{2}\nabla \chi -\frac{e}{c} {\bf A}.
\end{eqnarray}

In Ref.\ \onlinecite{14}, which is devoted to the general theory of
elasticity in superconductors, a relation is obtained between the fields
$\Phi $ and ${\bf p}_{s}$ and the elastic displacements ${\bf u}$. In the
case of longitudinal sound and an isotropic one-dimensional model, one can,
using Ref.\ \onlinecite{14}, write the following relations for the Fourier
components of $\Phi $ and $p_{s}$:
\begin{align}
&-a\Phi = cp_{sx} +i\omega c^{(d)} mu_x, \nonumber
\\
& -(ab+c^2)p_{sx} = i\omega[ab^{(d)}-cc^{(d)}] mu_x.
\end{align}

The solution of the boundary-value problem analogous to that considered above
but for superconductors meets with considerable difficulties due to the
energy dependence of the velocity of normal excitations,\cite{15} and
therefore for making estimates we limit consideration to the $q$ contribution
only, for which the wave vector of the Fourier component coincides with the
wave vector of the sound. Asymptotic expressions for the polarization
coefficients $a$, $b$, and $c$ of the electron subsystem and the
electroacoustic coefficients $c^{(d)}$ and $b^{(d)}$, which were all found in
Ref.\ \onlinecite{14} in the limit of strong spatial dispersion ($ql \gg 1)$,
have the form
\begin{align}
a&=1+\frac{i\pi}{2}\frac{s}{v_F} f(\Delta), \quad b = \frac{L}{m}\biggl[
\rho_s -\frac{3\pi i}{2} \frac{s}{v_F}
\frac{\Delta/4T}{\cosh^2(\Delta/4T)}\biggr], \nonumber
\\
c &=s(1-\rho_s), \quad b^{(d)} = [\frac{i\pi}{2}\frac{s}{v_F} \frac{L}{m}
\biggl( 1+\frac{\Delta}{2T} \ln\frac{Tv_F}{\Delta s} \biggr), \nonumber
\\
c^{(d)} &= -\frac{i\pi}{2} \frac{L}{mv_F} f(\Delta),
\end{align}
where $\rho _{s}\approx 2(T_{C}-T)/T_{C}$ is the density of the
superconducting condensate, $f(\Delta ) = 2/(\exp(\Delta /T)+1)$ is the
Fermi function, and $\Delta (T)$ is the superconducting gap.

Restricting consideration in (18) and (19) to the first nonvanishing terms of
the expansion in the small parameter $s/v_{F}$ and assuming that the
inequalities $\rho _{s},\Delta /T>(s/v_{F})^{2}$ hold, we find
\begin{align}
\Phi &= i\omega \frac{cb^{(d)} -bc^{(d)}}{ab+c^2} mu_x = -i\omega c^{(d)}
mu_x, \nonumber
\\
p_s &= -i\omega \frac{ab^{(d)}+cc^{(d)}}{ab+c^2} mu_x =
-i\omega\frac{b^{(d)}+sC^{(d)}}{b} mu_x.
\end{align}

In the case of longitudinal sound the vector potential in (17) can be
dropped, and we obtain for the electric potential
\begin{eqnarray}
e\varphi (x) = \Phi(x)-sp_s(x).
\end{eqnarray}

It follows from (19)--(21) that the contribution $p_{s}$ to the periodic
component of the potential is small in the parameter $s/v_{F}$ and can also
be dropped, and the relation between the amplitudes of the potential and the
normal ($n$) and superconducting ($sc$) states has the form
\begin{eqnarray}
(e\varphi)_{sc} = (e\varphi)_n f(\Delta),
\end{eqnarray}
i.e., the $q$ contribution should fall off with decreasing temperature
in the same way as the sound attenuation coefficient. In addition, it
follows from (22) that the phase of the potential should not change
on transition through $T_{C}$. Since the $K$ component of the
potential is also due to the deformation interaction, there is no
reason to think that its law of variation will be substantially
different from (22).

\begin{figure}
\centerline{\epsfxsize=8cm\epsffile{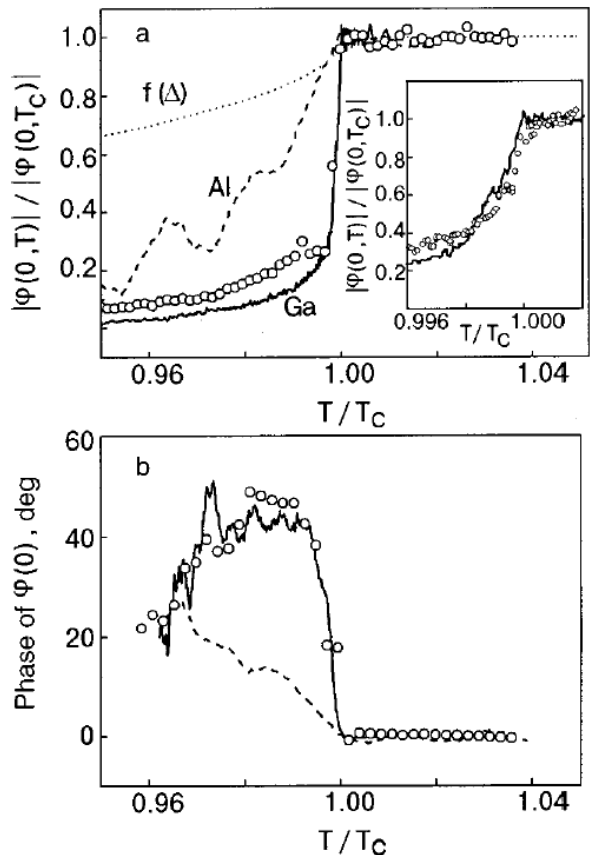}}\vspace{-5mm}
\caption{Effect of the superconducting transition on the electric potential:
amplitude $|\varphi (0)|$ --- the solid and dashed curves are for a galvanic
contact (Ga and Al), the circlets are for an asymmetric coil (Ga), and the
upper dotted curve shows $f(\Delta )$. Inset: the behavior of $|\varphi (0)|$
in Ga (galvanic contact) for different excitation amplitudes: solid curve ---
+10 dB, circlets --- 0 dB (a); the variation of the phase $\varphi (0)$ below
$T_{C}$: the solid and dashed curves are for a galvanic contact (Ga and Al),
and the circlets are for an asymmetric coil (b). } \label{f6}\vspace{-5mm}
\end{figure}

The experimental results presented in Fig.\ \ref{f6} strongly
contradict the estimates given above. Even in Al, in which apparently
only the periodic component of the signal is present to a significant
degree, the amplitude of $\varphi (0)$ changes much more sharply than
$f(\Delta )$, and the phase undergoes a rather rapid rise. These
features are more pronounced in Ga, in which one might suspect a
manifestation of some sort of nonlinearity, since a nonlinearity is expressed
quite clearly in the sound attenuation.\cite{16} For an excitation
intensity close to the maximum, nonlinear behavior of $\varphi (0)$ was
indeed observed, associated with both overheating and a sharp drop in
the sound attenuation coefficient below $T_{C}$, with a corresponding
decrease in the heat release. The data presented in Fig.\ \ref{f6}a
pertain to the region of amplitudes in which these effects are
practically absent (inset in Fig.\ \ref{f6}a).

The following ``simplified'' line of reasoning is also possible. In the
experiments described, the potential difference between the ``hot'' spot and
the remote parts of the sample is recorded in the regime of spreading surface
currents. Therefore, the value of the potential that is registered can depend
on the spreading resistance. In other words, in a superconductor the rapid
growth of the diamagnetic contribution to the conductivity (${\sim }\rho
_{s}v_{F}/s)$ can lead to ``short-circuiting'' of the source of the
electromotive force, and that is what is detected in the experiment. In Fig.\
\ref{f6} we also show the signal taken from the asymmetric coil, which
registered the amplitude and phase of the spreading currents specifically. We
see that the amplitude of this signal falls off just as rapidly as does
$|\varphi (0)|$, and the phase also increases practically in a jump.
Therefore, from the standpoint of the ``simplified'' explanation a rapid
growth of the ``resistance'' of the source of emf must also be assumed, and
that is hard to imagine.

From the results presented in Fig.\ \ref{f6}a for Ga we see that beyond
the region of very rapid drop of $|\varphi (0)|$ comes a more gradual
``tail,'' reminiscent of the evolution of $f(\Delta $). This suggests
another hypothesis: for some reason that we have not discussed, the $K$
contribution in a superconductor actually vanishes very rapidly, and
the remaining $q$ contribution falls off as follows from the theory. We
recall, however, that the phase of the $q$ contribution is different
from the phase of the $K$ contribution by more than $\pi /2$ (Fig.\
\ref{f3}b), and so this scenario should be accompanied by a downward
phase jump of corresponding amplitude. The data of Fig.\ \ref{f6}b are
inconsistent with this hypothesis as well.

At present the authors do not have any reasonable explanation for the
described behavior of $\varphi (0)$ near $T_{C}$. On the whole, the picture
looks as if either a state in which an uncompensated charge cannot exist
develops with catastrophic rapidity in a superconductor which is found in the
normal phase, at least in the nonlocal limit, or else below $T_{C}$ the
electric potential just as rapidly transforms to a quantity that cannot be
measured by a voltmeter. We add that a jump in the phase of $\varphi (0)$ by
approximately $\pi /4$ can be evidence that the surface impedance has some
relation to the observed effect, since the penetration depth of the field at
the transition through $T_{C}$ changes from a complex quantity with
approximately equal real and imaginary parts (their relative sizes depend on
whether the normal or anomalous skin-effect regime is realized) to the purely
real London penetration depth.

\section*{5. CONCLUSION}

In summary, we have carried out a theoretical analysis and experimental
observation and study of the electric potential arising when a longitudinally
polarized elastic wave is incident normally on a metal surface. The potential
under study is the sum of two contributions. The first of them, the $q$
contribution, is due to forced oscillations of the electric field, which is
proportional to the elastic deformation $u'$, i.e., it can be detected only
on a non-free boundary, where $u'(0)\neq 0$. The second, $K$ contribution is
due to the presence of the metal boundary itself, which distorts the
ballistic motion of nonequilibrium carriers in a subsurface layer of
thickness ${\sim }l$; its amplitude turns out to be proportional to the
displacement of the surface, $u(0)$. In the nonlocal limit the amplitude of
the $K$ contribution is markedly greater than that of the $q$ contribution,
while in the local limit it is the other way around. The phase of the $K$
contribution leads the phase of the $q$ contribution by approximately $\pi
/2$, and therefore in the crossover region the phase of the recorded
potential varies quite sharply. The experimental observation of just such
behavior is, in the opinion of the authors, an unambiguous qualitative
confirmation of the correctness of the ideas developed in this paper about
the origin of the potential in question.

At the superconducting transition the amplitude of the potential falls
in a catastrophically rapid manner, and the phase just as rapidly
increases by approximately $\pi /4$. This behavior is inconsistent with
the theoretical concepts, which predict a much smoother decrease of the
amplitude, close to the BCS dependence of the longitudinal sound
attenuation coefficient, and the absence of any phase variations.

The authors express their deep gratitude to E. A. Masalitin for invaluable
contribution to the development and preparation of the measuring apparatus,
which was ideally suited for addressing the problems investigated in this
study.


\begin{thebibliography}{99}
\bibitem{1a}
This is not strictly accurate. Applications of electroacoustic conversion
often make use of the circumstance that in a quasistatic electric field the
surface of a conductor is subjected to an additional pressure. In an
alternating field this pressure excites elastic displacements of the surface,
including longitudinal ones. This effect is widely used in low-frequency
acoustic experiments with vibrating reeds,\cite{2} and it is also the basis
of the ESAT (electrostatic acoustic transformation) mechanism.\cite{3}.Since
the response is quadratic in the field amplitude, in this case the
reciprocity principle is, of course, not obeyed, and this effect can be used
for acoustoelectric conversion only with additional ``linearization''. In the
present paper we discuss only linear effects.
\bibitem{15a}
We note that for the customary way of representing the elastic field [${\sim
}\exp(-i\omega t+i{\bf q}\cdot {\bf r})]$, in comparing the measured phase
variations with those calculated using relation (13), the sign of the latter
should be changed.
\end{thebibliography}
\end{document}